\documentclass{PoS}
\usepackage{amssymb,amsmath,wrapfig}

\title{Supersymmetry on the Lattice - Exact Results for Supersymmetric Quantum Mechanics}
\title{Exact results for supersymmetric
  quantum mechanics on the lattice}

\ShortTitle{Exact results for SUSY QM on the lattice}

\author{\speaker{David Baumgartner} and Urs Wenger\\
	Albert Einstein Center for Fundamental Physics\\
	Institute for Theoretical Physics\\
        University of Bern\\
	Sidlerstrasse 5\\
	CH-3012 Bern\\
	Switzerland\\
        E-mail: \email{baumgart@itp.unibe.ch,wenger@itp.unibe.ch}}

      \abstract{We discuss $\mathcal{N}=2$ supersymmetric quantum
        mechanics on the lattice using the fermion loop
        formulation. In this approach the system naturally
        decomposes into a bosonic and fermionic
        sector. This allows us to deal with the sign problem arising
        in the context of broken supersymmetry due to the vanishing of
        the Witten index. Employing transfer matrix techniques we
        obtain exact results at finite lattice spacing and are hence able
        to study how the continuum limit is approached. In particular,
        we determine how supersymmetry is restored and how, in the
        case of broken supersymmetry, the goldstino mode emerges.}

\FullConference{The XXIX International Symposium on Lattice Field Theory\\
                 July 10-16, 2011\\
                 Squaw Valley, California, USA}

\begin{document}

\section{Introduction}

Calculating nonperturbative properties of supersymmetric theories on
the lattice encounters various difficulties related to the fact that
the discretisation of space-time explicitly breaks supersymmetry
and violates Leibniz' rule. Moreover, the vanishing of the Witten
index in the context of spontaneous supersymmetry breaking leads to a
fermion sign problem which makes straightforward numerical simulations
impossible. While the restoration of supersymmetry can sometimes be
achieved in the continuum limit of the lattice theory, e.g.~by fine
tuning or by constructing $Q$-exact discretisations
\cite{Catterall:2010jh}, a solution to the sign problem is not easy to
find. A possible way out has been proposed in
\cite{Wenger:2008tq,Wenger:2009mi,Baumgartner:2011cm}. It is based on
the fermion loop formulation which can be simulated without critical
slowing down even when a massless goldstino mode is present.

In this work, we apply the fermion loop formulation to $\mathcal{N} =
2$ supersymmetric quantum mechanics for superpotentials yielding broken or
unbroken supersymmetry. Using transfer matrix techniques we are able
to obtain exact results for partition functions and various
observables at finite lattice spacing. We investigate how the
supersymmetric spectrum is recovered in the continuum limit and how
the goldstino mode emerges in the case of broken supersymmetry. In
these proceedings we confine ourselves to the presentation of results
obtained using a Wilson type discretisation together with the
appropriate fine tuning of counterterms, although results using a
$Q$-exact discretisation have been derived as well.

\section{Supersymmetric quantum mechanics on the lattice}
The continuum action of $\mathcal{N} = 2$ supersymmetric quantum
mechanics can be written as
\begin{equation}\label{eq:continuum_action}
 S = \int dt  \left[ \frac{1}{2}\left( \frac{d \phi(t)}{dt}\right)^2 +
   \frac{1}{2}P'(\phi(t))^2 +
   \overline{\psi}(t)\left(\frac{d}{dt}+P''(\phi(t))\right)\psi(t) \right]
\end{equation}
with one real bosonic coordinate $\phi$, two anticommuting fermionic
coordinates $\overline{\psi}$ and $\psi$, and a generic superpotential
$P(\phi)$. The derivative of the superpotential $P'(\phi)$ is taken
with respect to $\phi$, $P'(\phi) \ \dot{=} \ \frac{\partial
  P(\phi)}{\partial \phi}$. For periodic boundary conditions ($PBC$)
the action is invariant under two supersymmetry transformations
$\delta_{1,2}$:
\begin{displaymath}
\begin{array}{rclcrcl}
\delta_1 \phi &=& \psi \overline{\epsilon}, &\quad& \delta_2 \phi &=& \overline{\psi} \epsilon, \\ 
\delta_1 \psi &=& 0, &\quad&  \delta_2 \psi &=&  (\frac{d\phi}{dt} - P') \epsilon,\\
\delta_1 \overline{\psi} &=& (\frac{d\phi}{dt} + P')  \overline{\epsilon}, &\quad&  \delta_2 \overline{\psi} &=& 0,
\end{array}
\end{displaymath}
with two Grassmann valued parameters $\epsilon$ and
$\overline{\epsilon}$. Note that for supersymmetric quantum mechanics
it is the form of the superpotential $P(\phi)$ which determines the
supersymmetry breaking pattern.  If the highest power of $P(\phi)$ is
even (odd), supersymmetry is unbroken (broken).  A main feature of
supersymmetry is the degeneracy between the energy levels in the
bosonic and the fermionic sector. For unbroken supersymmetry, however,
there is one single unpaired energy level at zero energy, i.e., a
unique ground state, either in the bosonic or in the fermionic
sector. This is in contrast to the case of broken supersymmetry, where
the lowest energy levels in both sectors are degenerate and lifted
above zero.  In addition, there is a zero energy goldstino mode which
mediates between the two degenerate ground states.

The supersymmetry breaking pattern can also be partly infered from the
Witten index. It is formally defined as
\begin{displaymath}
 W \equiv \lim_{\beta \rightarrow \infty} \mathrm{Tr} \left[ (-1)^F
   \exp(-\beta H) \right],
\end{displaymath}
where $F$ denotes the fermion number operator and $H$ is the
Hamiltonian of the system. Essentially, $W$ counts the difference
between the number of bosonic and fermionic zero energy states and its
vanishing provides a neccessary but not sufficient condition for
supersymmetry breaking. $W$ can also be written more explicitely as
\begin{equation}
\label{eq:def_witten_index}
 W = \lim_{\beta \rightarrow \infty} [\mathrm{Tr}_b \exp(-\beta
 H) - \mathrm{Tr}_f \exp(-\beta H)] = \lim_{\beta \rightarrow
   \infty}[Z_0 - Z_1] = \lim_{\beta \rightarrow \infty}Z_{PBC} \, ,
\end{equation}
where $\mathrm{Tr}_{b,f}$ denote the traces over the bosonic and
fermionic states. $Z_{0,1}$ are the partition functions in the
$F=0,1$ sectors and $Z_{PBC}$ is the one with periodic boundary
conditions.  In the language of field theory, the latter can be
calculated via
\begin{displaymath}
 Z_{PBC} = \int \mathcal{D}\phi \mathcal{D}\overline\psi
 \mathcal{D}\psi \exp(-S) = \int_{-\infty}^\infty \mathcal{D}\phi \det D(\phi) \exp(-S_\phi).
\end{displaymath}
In the last step, the fermions have been integrated out yielding the
fermion matrix determinant $\det D$ and the bosonic part of the
action $S_\phi$. In this representation, the origin of a fermion sign
problem becomes evident when supersymmetry is broken: a vanishing Witten
index requires the determinant $\det D$ to be indefinite.

\subsection{Lattice formulation}
For the construction of a lattice version of the model, we follow
Golterman and Petcher \cite{Golterman:1988ta} and employ the same
lattice derivative for the bosons as for the fermions. To avoid
fermion doublers, we use the Wilson lattice derivative with Wilson
parameter $r = 1$. In one dimension this simplifies to the backward
derivative $(\Delta^-f)_x = f_x - f_{x-1}$ and the discretised action
explicitly reads
\begin{equation}\label{stand_lat_action}
 S_L = \sum_x \left[ \frac{1}{2}(P'(\phi_x)^2 + 2\phi_x^2) -
   \phi_x\phi_{x-1} + (1 + P''(\phi_x))\overline{\psi}_x\psi_x -
   \overline{\psi}_x\psi_{x-1} \right].
\end{equation}
Due to radiative corrections the lattice theory is, however, not
guaranteed to yield a supersymmetric theory in the continuum
limit. The corrections can be accounted for either by adding a
suitable counterterm $\frac{1}{2}\sum P''$ to the action
\cite{Golterman:1988ta,Giedt:2004vb}, which restores the
supersymmetries in the continuum limit, or by adding the surface term
$\sum P'(\Delta^-\phi)$
\cite{Catterall:2000rv,Catterall:2003wd,Bergner:2007pu} resulting in a
$Q$-exact action. The latter construction preserves a particular combination
of the supersymmetries $\delta_{1,2}$ exactly even at finite lattice
spacing and hence guarantees the correct continuum limit without any
fine tuning.

To circumvent the sign problem discussed above, we make use of the
fermion loop formulation
\cite{Wenger:2008tq,Wenger:2009mi,Baumgartner:2011cm}. The basic idea
here is to exactly rewrite the exponential of the fermion degrees of
freedom as a power series to all orders. Upon integration of the
fermion fields, the nilpotency of the Grassman variables yields a
constraint on the oriented fermionic bond occupation numbers $n_x^f =
0, 1$ between the sites $x$ and $x-1$ related to the fermion hopping
term $\overline \psi_x \psi_{x-1}$, and on the monomer occupation
numbers $m_x^f = 0, 1$ stemming from the term $(1 +
P''(\phi_x))\overline{\psi}_x \psi_x$. The constraint is given by
\begin{displaymath}
m_x^f + \frac{1}{2}(n_x^f + n_{x + 1}^f) = 1 \quad \forall x,
\end{displaymath}
and allows only two fermion configurations: $\{m_x^f=1, n_x^f=0, \,
\forall x\}$ with fermion number $F=0$, and $\{m_x^f=0, n_x^f=1, \,
\forall x\}$ with fermion number $F=1$. For PBC the latter receives an
additional minus sign relative to the former due to the fermion
loop. As a consequence, the partition function naturally decomposes
into a bosonic and fermionic contribution $Z_0$ and $Z_1$, in
accordance with eq.(\ref{eq:def_witten_index}). It is this
decomposition which eventually allows to take care of the fermion sign
problem.

In addition to the fermion bonds and monomers we also introduce
non-oriented bonds for the bosonic degrees of freedom, with the
corresponding bosonic bond occupation numbers $n_x^b \in \mathbb{N}^0$
\cite{Baumgartner:2011cm}. Depending on the symmetries of the action,
these bosonic bond configurations may obey certain constraints. By
summing over these constrained configurations $\{n_x^b\}$ we obtain
the locally factorised partition functions with fixed fermion number
$F=0,1$,
\begin{displaymath}
  Z_F  = \sum_{\{n_x^b\}} \prod_x \frac{1}{n_x^b!} Q_F(N_x)
\end{displaymath}
where the local weights $Q_F$ are defined as
\begin{displaymath}
 Q_F(N) = \int d \phi \ \phi^N \mathrm{e}^{-\frac{1}{2}(P'(\phi)^2+2\phi^2)}(1+P''(\phi))^{1-F} 
\end{displaymath}  
with the bosonic site occupation number $N_x = n_x^b + n_{x+1}^b$. The
$Q$-exact discretisation requires additional types of bosonic bonds,
but still leads to a locally factorised partition function.

\subsection{Transfer matrix}
The dimensionality of the system allows a further reformulation in
terms of a transfer matrix between states defined on the dual
lattice. Each state is characterised by the fermion bond occupation
number number $n^f$ and the boson bond occupation number $n^b$,
i.e.~$|n^f,n^b \rangle$. Since the fermion number is conserved the
transfer matrix has a block structure consisting of the two matrices
$T^{F=0,1}_{m^b,n^b}$ which take the system from state $|F,n^b\rangle$
to $|F,m^b\rangle$. To be specific, the transfer matrix elements are
given by
\begin{displaymath}
 T^F_{m^b,n^b} = \frac{1}{\sqrt{m^b!}} \frac{1}{\sqrt{n^b!}} Q_F(m^b + n^b)\,.
\end{displaymath}
In order to keep the size of the matrices finite, we introduce a
cutoff on the maximal bosonic bond occupation number. Keeping it of
the order ${\cal O}(10^2)$ turns out to be sufficient to render all results
independent of the cutoff.

In terms of these transfer matrices, the partition function for a
system with $L_t$ lattice sites is calculated in each sector $F$
according to
\begin{displaymath}
 Z_F = \mathrm{Tr}[(T^F)^{L_t}].
\end{displaymath}
These partition functions can then be combined to $Z_{PBC} = Z_0 -
Z_1$ and $Z_{aPBC} = Z_0 + Z_1$ for $PBC$ and antiperiodic boundary
conditions ($aPBC$), respectively.  The construction via the transfer
matrices allows the straightforward calculation of various
observables, such as correlation functions, Ward identities and mass
gaps.  The latter are directly associated with the eigenvalues of the
transfer matrices. If we denote the eigenvalues of $T^F$ by
\mbox{$\lambda^F_0>\lambda^F_1>\dots$}, the $k$-th bosonic mass gap in
the sector $F$ can be calculated as
\begin{displaymath}
m_b^{F,k} = -L_t \cdot \log \left(
     \lambda^F_k/\lambda^F_0 \right), \quad k=1,2,\ldots\,,
\end{displaymath}
whereas the $k$-th fermionic energy gap is given by
\begin{displaymath}
m_f^{F,k} = -L_t \cdot \log \left( \lambda^{1-F}_k/\lambda^F_0 \right) , \quad k=0,1,\ldots\, .
\end{displaymath}
The transfer matrix approach can, of course, be generalised
straightforwardly to any kind of discretisation of the action
eq.(\ref{eq:continuum_action}), in particular also to the $Q$-exact
discretisation.

\section{Results}
We now present the results for the action (\ref{stand_lat_action})
with the counterterm using the techniques introduced above.  For our
calculations, we use the superpotential \mbox{$P_u(\phi) =
  \frac{1}{2}m\phi^2 + \frac{1}{4}g\phi^4$} as an example with
unbroken supersymmetry and \mbox{$P_b(\phi) =
  -\frac{m^2}{4\lambda}\phi + \frac{1}{3}\lambda\phi^3$} as an example
for which the supersymmetry is broken.  The calculations are performed
at coupling strengths $g/m^2 = 1.0$ and $\lambda/m^{3/2} = 1.0$,
respectively, thus we are clearly in a regime where perturbation
theory is not applicable. For a system with aPBC for the fermion the
temporal extent of the lattice is inversely related to the temperature
$T$ of the system, such that $mL \rightarrow \infty$ corresponds to
the zero temperature limit. Finally, the continuum limit is reached by
taking $L_t \rightarrow \infty$.

\subsection{Witten index}
\begin{figure}[t!]
\centering
\includegraphics[width=0.49\textwidth,angle=00]{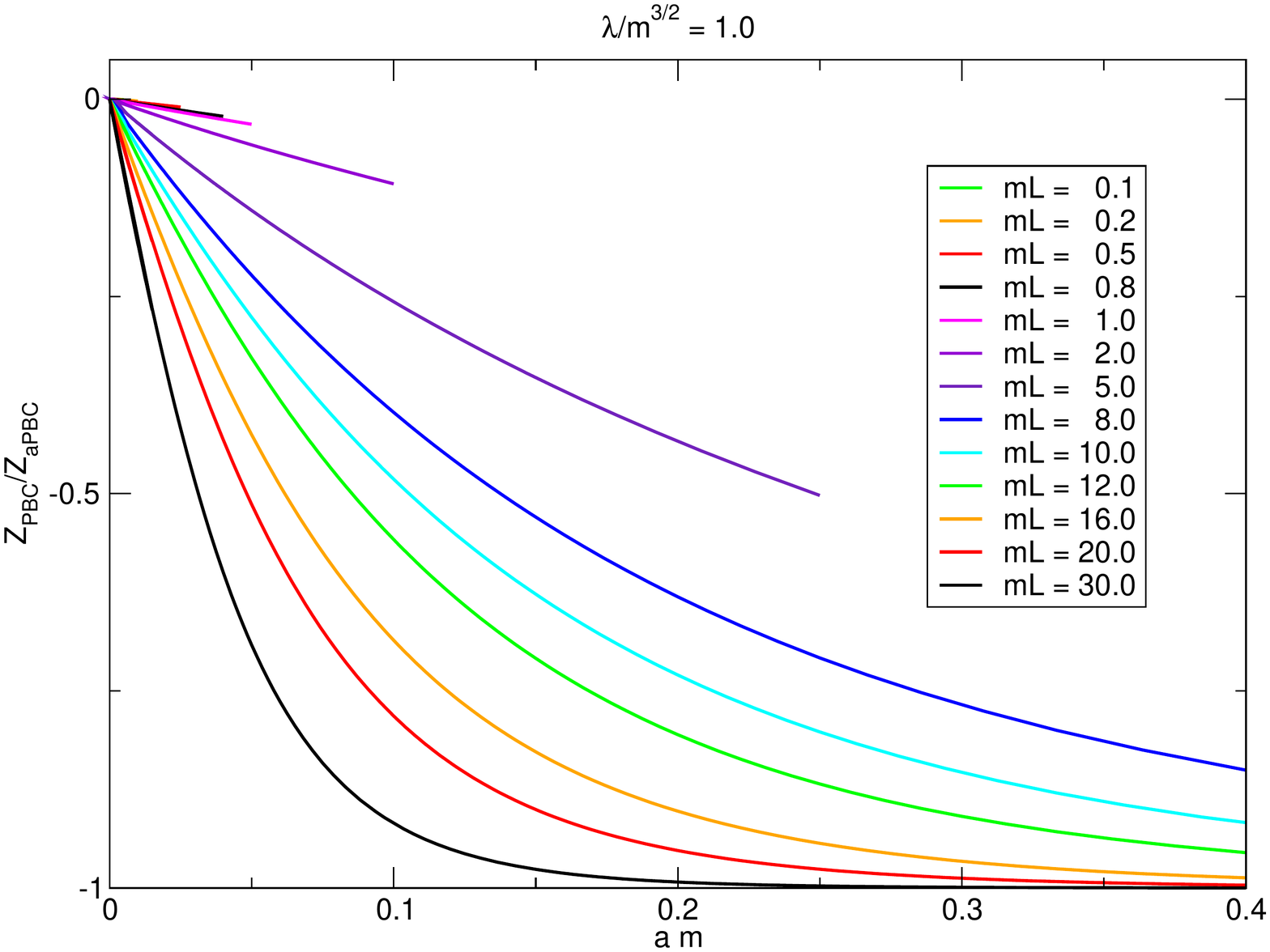}
\includegraphics[width=0.49\textwidth,angle=00]{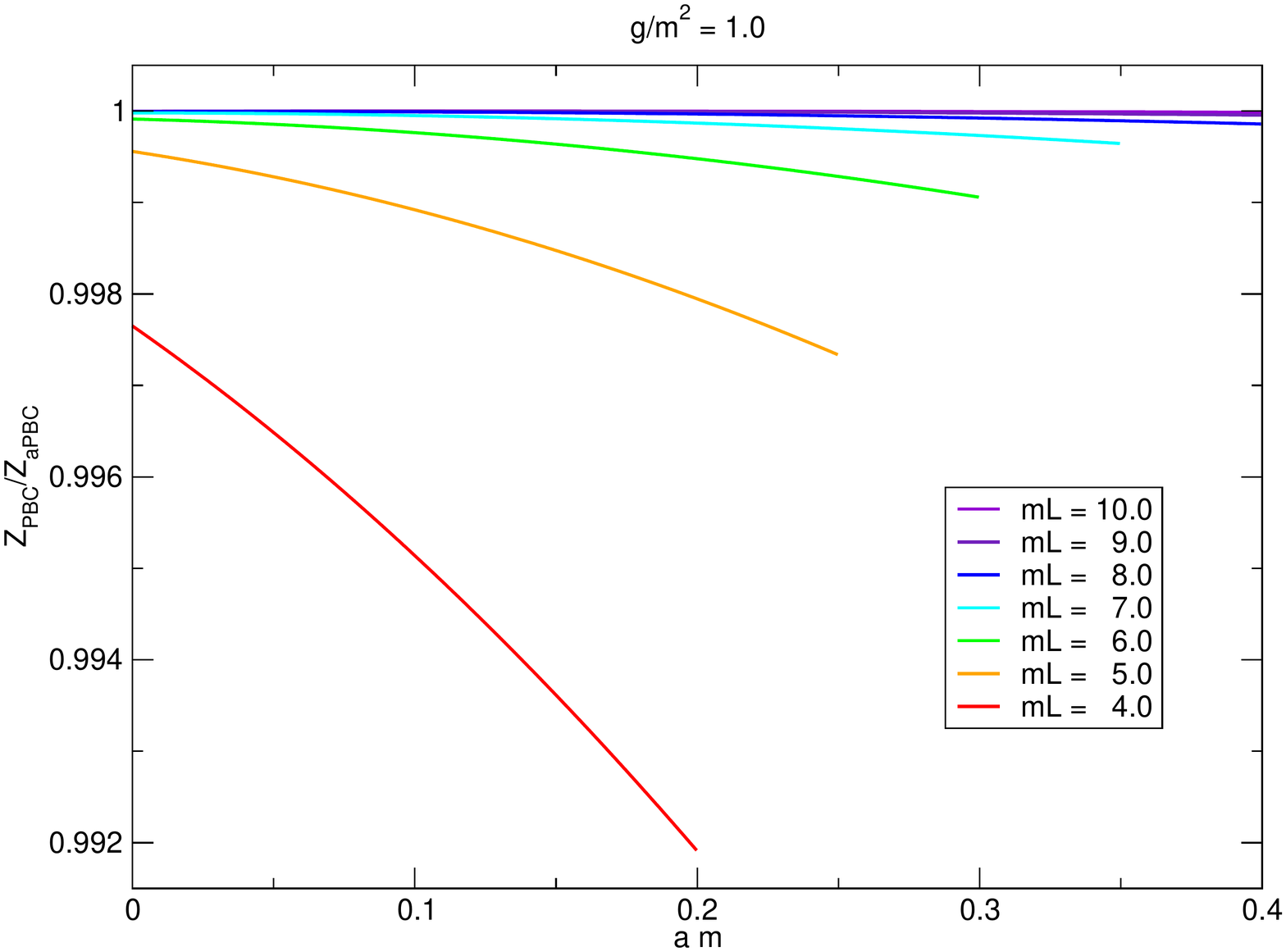}
\caption{The Witten index $W = Z_{PBC}/Z_{aPBC}$ versus the lattice
  spacing for broken (left plot) and unbroken supersymmetry (right
  plot) at various values of the inverse temperature $mL$. The
  continuum limit corresponds to $ am \rightarrow 0$.  }
\label{fig:WittenIndex}
\centering
\end{figure}

The Witten index is determined by the quantity $Z_{PBC}/Z_{aPBC} =
(Z_0-Z_1)/(Z_0+Z_1)$. It measures the relative weight between the
bosonic and fermionic sectors $Z_0$ and $Z_1$, respectively. In the
system with broken supersymmetry both ground states are equally
favourable, yielding $Z_{PBC}/Z_{aPBC} = 0$ in the zero temperature
limit. Of course, the degeneracy between the two ground states is
broken at finite lattice spacing, so one expects a Witten index $W=\pm
1$ in the limit $T\rightarrow 0$ at fixed $a$. It turns out that for
our choice of parameters, the fermionic ground state has a slightly
lower energy at finite $a$ leading to $W=-1$ in the $T\rightarrow 0$
limit, {\it cf.}~left plot in figure \ref{fig:WittenIndex}. This is
true for
any finite $a$, so the order of the limits $\lim_{T\rightarrow 0}
\lim_{a\rightarrow 0}$ is crucial to obtain $W=0$. Note also that for
$T \gg 1$ the Witten index tends to zero at any finite $a$, in
accordance with the counting of the states in
eq.(\ref{eq:def_witten_index}) at finite temperature.

In the situation with unbroken supersymmetry the system is forced to
occupy the single unique ground state in the zero temperature limit,
yielding $Z_{PBC}/Z_{aPBC} = +1$ or $-1$. Note that for our specific
choice of parameters the ground state is bosonic, hence
$W=+1$. However, while one finds that the index is pushed away from 0
for $T \gg 1$ as before, in the limit $T\rightarrow 0$ it will always
go to 1 at any finite $a$, {\it cf.}~right plot in figure
\ref{fig:WittenIndex}. So it turns out that for unbroken symmetry, the
order of the two limits $\lim_{T\rightarrow 0}$ and $\lim_{a\rightarrow 0}$ is not
relevant.

\subsection{Mass gaps}
\begin{figure}[t!]
\centering
\includegraphics[width=0.49\textwidth,angle=00]{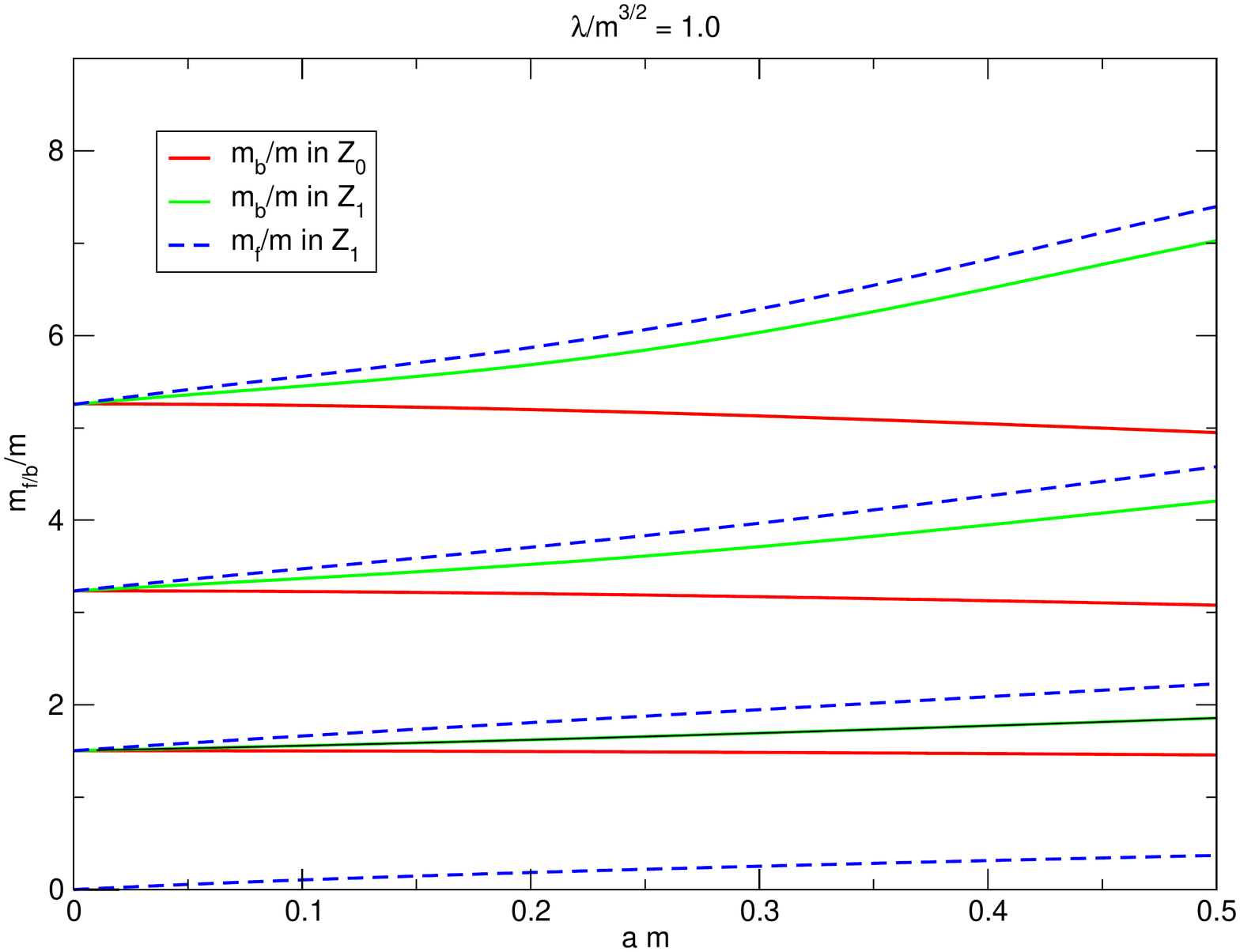}
\includegraphics[width=0.49\textwidth,angle=00]{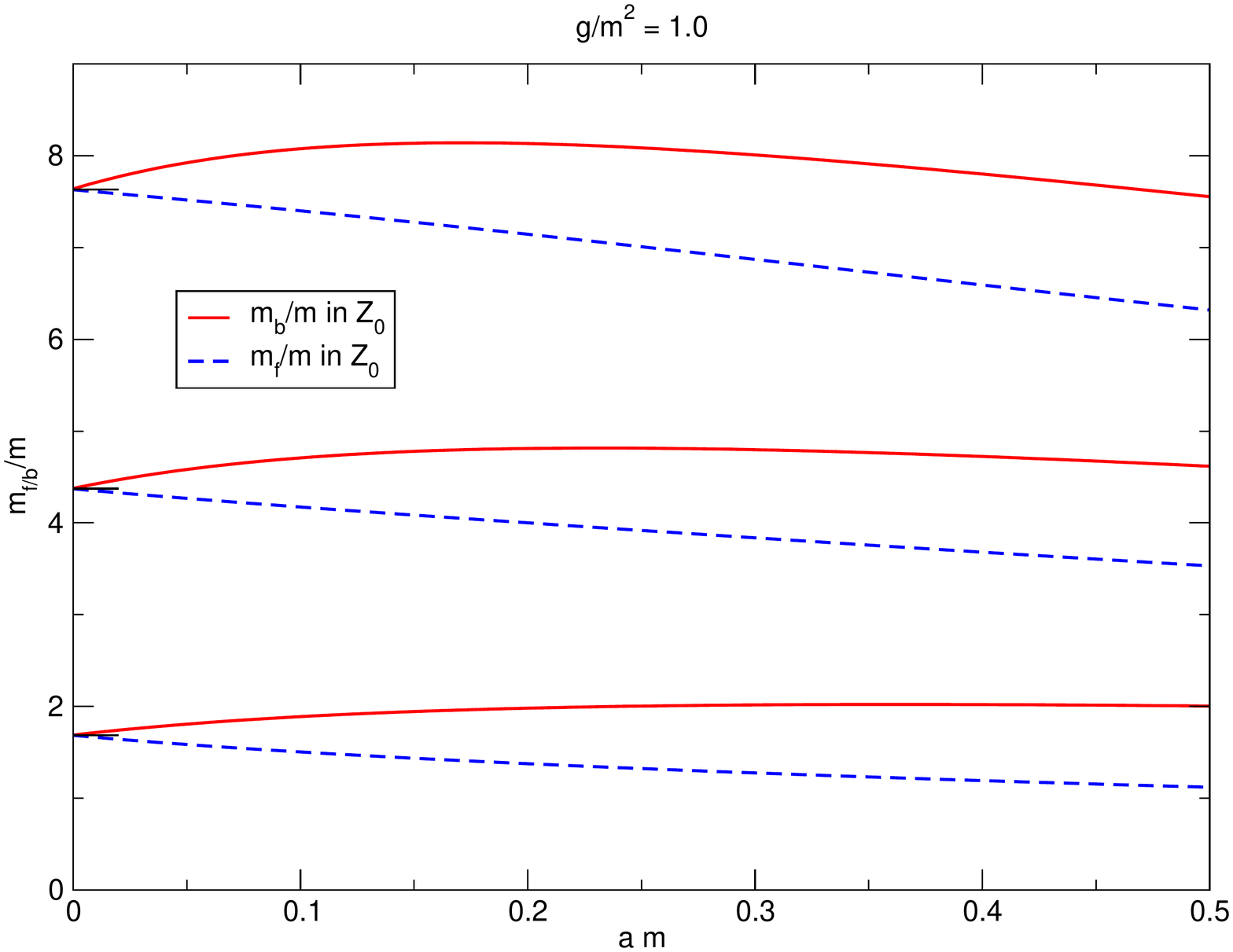}
\caption{Bosonic and fermionic mass gaps $m_b$ and $m_f$ versus the
  lattice spacing $a$, all expressed in units of the bare mass
  $m$. Results for broken and unbroken supersymmetry are displayed in
  the left and right panel, respectively.  Note that for broken
  supersymmetry the zero energy Goldstino mode emerges in the
  continuum limit.  }
\label{fig:EnergyGaps}
\centering
\end{figure}

It is also interesting to study how the energy or mass gaps approach
the continuum limit. In figure \ref{fig:EnergyGaps} we show the
results for the lowest few masses as a function of the lattice spacing
$a$, everything expressed in units of the bare mass $m$, for broken
(left plot) and for unbroken supersymmetry (right plot). Since we
extract the mass gaps from the eigenvalues of the transfer matrices,
the results are obtained directly in the limit $T\rightarrow 0$. As a
consequence, for broken supersymmetry, where there are two degenerate
ground states in the continuum, it makes sense to calculate bosonic
excitations $m_b$ both in the $F=0$ and $F=1$ sector. The plots
illustrate nicely how the supersymmetry in the spectrum, i.e.~the
degeneracies between the bosonic and fermionic excitations, are
restored in the continuum limit. Furthermore, when the supersymmetry
is broken one expects a zero energy fermionic excitation, the
goldstino mode, which is responsible for the fact that $Z_{PBC} =
0$. From the plot it becomes clear how the lattice acts as a regulator
for the goldstino mode and, as a consequence, also for the vanishing
Witten index $W$, hence allowing to give meaning to (finite)
observables even in the system with PBC.  Finally, we make the
observation that the leading lattice artefacts of the spectral mass
gaps are all ${\cal O}(a)$ except for $m_b$ in the $F=0$ sector when
the supersymmetry is broken. In that case they are ${\cal O}(a^2)$.

\section{Conclusions}
We have presented exact results for $\mathcal{N} = 2$ supersymmetric
quantum mechanics on the lattice using the fermion loop formulation
and corresponding transfer matrices. With these techniques we are able
to study in detail how the supersymmetric spectrum is recovered in the
continuum limit and how the Witten index is regularised on the
lattice.

In the loop formulation the partition function naturally separates
into bosonic and fermionic contributions and this is crucial for
containing the fermion sign problem in supersymmetric systems with
broken supersymmetry. The transitions between the bosonic and
fermionic sectors are controlled by the (would-be) goldstino mode
which becomes massless only in the continuum limit. Since massless
fermion modes can be efficiently simulated with the fermion loop
algorithm proposed in \cite{Wenger:2008tq,Wenger:2009mi} our approach
provides a way to circumvent the sign problem. Indeed, results from
Monte Carlo simulations of $\mathcal{N} = 2$ supersymmetric quantum
mechanics have already been presented in \cite{Baumgartner:2011cm} and
in this work we have provided the corresponding exact results using
transfer matrices.

It is also interesting to apply our approach in higher dimensions
where it allows to investigate the spontaneous breaking of
supersymmetry nonperturbatively and from first principles. In
particular, the approach can be applied to supersymmetric Wess-Zumino
models \cite{Baumgartner:2011cm} in $d=2$ dimensions and first results
from simulations of the $\mathcal{N} = 1$ model including one Majorana
fermion and one scalar field have been presented at this conference
\cite{Baumgartner:2011jw}.

\providecommand{\href}[2]{#2}\begingroup\raggedright\endgroup

\end{document}